\documentclass[final,3p,times,twocolumn]{elsarticle}
\DeclareTextSymbol{\degre}{T1}{6}
\DeclareTextSymbol{\degre}{OT1}{23}

\usepackage{graphics}
\usepackage{amsmath}

\usepackage{amssymb}

\usepackage{lineno}

\journal{Astrobiology}

\begin{document}

\begin{frontmatter}

\title{On the abundances of noble and biologically relevant gases in Lake Vostok, Antarctica}

\author{Olivier Mousis}
\ead{olivier.mousis@obs-besancon.fr}
\address{Universit{\'e} de Franche-Comt{\'e}, Institut UTINAM, CNRS/INSU, UMR 6213, Besan\c con Cedex, France}
\address{Universit\'e de Toulouse; UPS-OMP; CNRS-INSU; IRAP; 14 Avenue Edouard Belin, 31400 Toulouse, France}
\author{Azzedine Lakhlifi}
\address{Universit{\'e} de Franche-Comt{\'e}, Institut UTINAM, CNRS/INSU, UMR 6213, Besan\c con Cedex, France}
\author{Sylvain Picaud}
\address{Universit{\'e} de Franche-Comt{\'e}, Institut UTINAM, CNRS/INSU, UMR 6213, Besan\c con Cedex, France}
\author{Matthew Pasek}
\address{Department of Geology, University of South Florida, Tampa FL 33620, USA}
\author{Eric Chassefi{\`e}re}
\address{Univ. Paris-Sud, Laboratoire IDES, UMR 8148, Orsay, F-91405, France  / CNRS, Orsay, F-91405, France}

\begin{abstract}
Motivated by the possibility of comparing theoretical predictions of Lake Vostok's composition with future in situ measurements, we investigate the composition of clathrates that are expected to form in this environment from the air supplied to the lake by melting ice. In order to establish the best possible correlation between the lake water composition with that of air clathrates formed in situ, we use a statistical thermodynamic model based on the description of the guest-clathrate interaction by a spherically averaged Kihara potential with a nominal set of potential parameters. We determine the fugacities of the different volatiles present in the lake by defining a ``pseudo'' pure substance dissolved in water owning the average properties of the mixture and by using the Redlich-Kwong equation of state to mimic its thermodynamic behavior. Irrespective of the clathrate structure considered in our model, we find that xenon and krypton are strongly impoverished in the lake water (a ratio in the 0.04--0.1 range for xenon and a ratio in the $\sim$0.15--0.3 range for krypton), compared to their atmospheric abundances. Argon and methane are also found depleted in the Lake Vostok water by factors in the 0.5--0.95 and 0.3--0.5 ranges respectively, compared to their atmospheric abundances. On the other hand, the carbone dioxide abundance is found substantially enriched in the lake water compared to its atmospheric abundance (by a factor in the 1.6--5 range at 200 residence times). The comparison of our predictions of the CO$_2$ and CH$_4$ mole fractions in Lake Vostok with future in situ measurements will allow disentangling between the possible supply sources.
\end{abstract}


\end{frontmatter}

\linenumbers

\section{Introduction}
\label{intro}

Over 100 liquid water lakes have been identified beneath the ice sheets of Antarctica (Smith et al. 2009). Among them, Lake Vostok lying buried beneath $\sim$4 kilometers of ice is the largest known subglacial lake on Earth. Estimates for the age of this lake range from one million years (Kapitsa et al. 1996) to 15 million years (Bell et al. 2002). Because Lake Vostok is under a very thick ice sheet, it is unlikely that gases within the lake water are in equilibrium with air. From a biological point of view, the study of Lake Vostok is interesting because it provides an environment that has been sealed off from light and atmosphere for possibly several millions of years. As such, the Lake Vostok environment could be analogous to that of the Jovian icy moon Europa, which is envisaged to harbor a subsurface ocean with conditions compatible with habitability (Hand et al. 2006; Pasek \& Greenberg 2012).

After cumulated decades of effort, researchers have succeeded in drilling through 4 kilometers of ice to the surface of the subglacial Lake Vostok on 5 February 2012 (Showstack 2012). The precise sampling of Lake Vostok in a near future will bring important constraints on the delivery processes of volatiles to the lake and the chemical pathways that may, or may not make life possible deep below the ice sheet.

Ice overlying the lake is characterized by air bubbles trapped at the air/ice interface that are compressed with increasing depth. As a consequence, these air bubbles transform to clathrate hydrates (hereafter clathrates) below a certain depth, depending on the temperature of the ice sheet. In the contact region between the bottom of the ice sheet and the lake surface, most of the air in the ice sheet is trapped within these clathrates. In situ measurements have shown that the average N$_2$/O$_2$ ratio of the clathrates in this contact region approaches the atmospheric value (Ikeda et al. 1999). Melting of these clathrates is thus responsible for a transfer of most of the atmospheric air to the Lake Vostok water through the ice sheet (Lipenkov \& Istomin 2001). However, in the contact region, clathrates could also be stable in the lake water thus reducing the amount of gases in aqueous solution. This might have an impact on the oxygen availability for possible life in the lake and it is thus very important to thoroughly investigate the composition of Lake Vostok.

The lake composition was previously investigated by McKay et al. (2003) (hereafter MK03). Based i) on the prediction that air clathrate is stable at depth higher than $\sim$1500 m in sub-glacial lakes (Lipenkov \& Istomin 2001) and ii) on the fact that selective gas trapping occurs during the formation of clathrates (van der Waals \& Platteeuw 1959; Miller 1974; Lunine \& Stevenson 1985), MK03 estimated the level of gases dissolved in Lake Vostok as a function of its evolution. To do so, they modeled the balance between the gas (i.e. air) supplied by ice melting and gas trapped by clathrate, and assuming that the lake is a closed system, with no gas escape. However, the approach followed by MK03 to calculate the relative partitioning of volatiles trapped in clathrate is not adapted for a gas mixture in which more than two species are included (Lunine \& Stevenson 1985), as it is the case for air clathrate. Also, the formalism used by the authors is a crude approximation of the statistical thermodynamic model that is usually used to investigate the composition of multiple guest clathrates. Indeed, it is based on the assumption that the Langmuir constants (see Section \ref{thermo}) are very close to each other for the different gases, irrespective of the considered type of clathrate cages (Lunine \& Stevenson 1985).

In this paper, motivated by the possibility of comparing theoretical predictions of Lake Vostok composition with in situ measurements in the near future, we reinvestigate the assumptions of MK03 by considering a more complete set of species dissolved in the lake. We thus determine more accurately the composition of clathrates that should form in this environment. In order to establish a better correlation between the lake composition with that of air clathrates formed in situ, we use a statistical thermodynamic model based on experimental data and derived from the original work of van der Waals \& Platteeuw (1959). This approach, used today in industry and science, has saved substantial experimental effort for the determination of: (i) the equilibrium pressure of a clathrate formed from various mixtures; and (ii) the mole fraction of the different species trapped in the clathrate from a given fluid phase (see Sloan \& Koh (2008) for details). The major ingredient of our model is the description of the guest-clathrate interaction by a spherically averaged Kihara potential with a nominal set of potential parameters. Section \ref{model} is devoted to the description of i) the strategy utilized to compute the lake composition and ii) the statistical thermodynamic model used in this work. In Section \ref{results}, we present the results of our calculations concerning the time evolution of clathrate and lake compositions. In Section \ref{discuss}, we compare our results with those of MK03 and discuss their implications for the physical and biological processes that may take place in Lake Vostok and analogous environments such as the putative internal ocean of the Jovian icy moon Europa.

\section{Modeling the lake composition}
\label{model}

\subsection{Computational approach}

Our basic assumptions are similar to those formulated by MK03. We assume that Lake Vostok is a closed system at constant pressure of 35 MPa with a mean temperature of -3$\degre$C, and is physiographically stable over time (Lipenkov \& Istomin 2001) for a description of Lake Vostok's thermodynamic conditions). In our system, water and air are delivered to the lake when melting occurs due to the slow downward motion of the overlying gas-rich ice layers, and gas-free water leaves the lake as ice accretes to the bottom of the ice sheet in regions where ice moves outward (Jouzel et al. 1999; Bell et al. 2002; MK03). Air is assumed to be supplied to the lake at a concentration of about 90 cm$^3$ at STP ($T=273.15$~K and $P = 0.1013$~MPa) per kg of melted ice (Jouzel et al. 1999; Lipenkov \& Istomin 2001). We express the age of the lake in units of the residence time (denoted by RT in the following), defined as the time required for all water of the lake to be renewed through the melting-freezing cycle of water flowing through the volume of the lake. Estimates of the RT for Lake Vostok range from 5,000 years (Philippe et al. 2001) to 125,000 years (Siegert et al. 2003). 
 
Because the clathrate formed in Lake Vostok is crystallized from dissolved air, it is expected to be dominated by N$_2$. A good approximation of its dissociation pressure is to assume it is equivalent to that of pure N$_2$ clathrate. The dissociation pressure $P^{diss}_{N_2}$ of N$_2$ clathrate follows an Arrhenius law (Miller 1961) defined as

\begin{equation}
log (P^{diss}_{N_2}) = A + \frac{B}{T},
\label{eq1}
\end{equation}

\noindent where $P^{diss}_{N_2}$ is expressed in Pa and $T$ is the temperature in K. The constants A and B are set to 9.86 and -728.58, respectively, and fit to experimental data (Thomas et al. 2007). At 270.15 K, $P^{diss}_{N_2}$ is found to be $\sim$14.6 MPa and the corresponding fugacity is $\sim$ 13.7 MPa (see Sec. \ref{fug}). However, at high hydrostatic pressure (here 35 MPa), the calculated dissociation fugacity needs a correction due to the specific volume difference between clathrate and the liquid solution. The effective dissociation fugacity of clathrate is found equal to $\sim$21.2 MPa (see Eq. \ref{eq1f}), which translates into an equivalent gas pressure of $\sim$22.5 MPa. The pressure of dissolved N$_2$ gas must increase with time up to this value to enable the triggering of clathrate formation in Lake Vostok. After having computed the Henry's constant of N$_2$ at the hydrostatic pressure of 35 MPa (see Sec. \ref{fug}), we find that the gas pressure of 22.5 MPa corresponds to a volume of $\sim$1735 cm$^3$ (STP) of N$_2$ per kg of water. These estimates imply that, at $\sim$25 RT, the concentration of air dissolved in Lake Vostok reaches a solubility limit of $\sim$2220 cm$^3$ (STP) per kg of water, a value lower than the one found by MK03 (2500 cm$^3$ at STP), just before the beginning of clathrate formation. 

Since the time for dissolved gases to diffuse through the lake is smaller than 1 RT, and due to the long timescales involved (several tens of RT, see hereafter), the gas may be supposed to be uniformly mixed in the lake water at 25 RT. Thereafter, the total amount of dissolved gas in Lake Vostok remains approximately constant and excess gas continuously forms clathrate. Because the composition of clathrate may depart from that of the coexisting gas phase, progressive clathrate formation could in turn influence the composition of the gas dissolved in Lake Vostok. Table \ref{airgas} gives the composition of the air dissolved in the lake water at the beginning of our computation (i.e. at 0 RT).

\begin{table*}
\centering
\caption{Gas dissolved in Lake Vostok. From left to right: composition of dry air by volume (Lodders \& Fegley 1998), critical temperatures and pressures (Lide 2002), partial molar volumes ($^a$Rudakov et al. 1996; $^b$Anderson 2002; $^c$Kennan \& Pollack 1990), Henry's constants at STP (Sander 1999) and calculated at 35 MPa of hydrostatic pressure and 270.15 K, and fugacities calculated at 25 RT and at pressures of 1 atm and 35 MPa, respectively. Properties of  species that are not incorporated in clathrate are intentionally left blank (see text).}
\rotatebox{0}{
\scalebox{0.9}{
\begin{tabular}{lllllllll}
\hline \hline
Species $K$ 	& $x_{K}^{air}$				& $T_c$		& $P_c$			& $\overline{V}_K$		& $H_{1atm,K}$		& $H_{hydro,K}$		& $f_{1atm,K}$				& $f_{hydro,K}$				\\
 			& 						& (K)			& (MPa)			& (cm$^3$/mol)		& (L atm mol$^{-1}$)		& (L atm mol$^{-1}$)			& (MPa)					& (MPa)					\\  
\hline
N$_2$   		& 0.7808					& 126.19		& 3.40			& 36$^a$				& 1639.34				& 2868.19					& 11.80					& 18.29					\\
O$_2$       	& 0.2095 					& 154.48		& 5.04			& 33$^a$				& 769.23				& 1284.54					& 1.49					& 2.30					\\
Ar               	& 9.34 $\times$ 10$^{-3}$  	& 150.87		& 4.90			& 33$^a$				& 714.29				& 1192.79					& 6.15 $\times$ 10$^{-2}$		& 9.53 $\times$ 10$^{-2}$		\\
CO$_2$		& 3.50 $\times$ 10$^{-4}$		& 304.25		& 7.38			& 38.4$^b$			& 28.57				& 51.88					& 9.22 $\times$ 10$^{-5}$		& 1.43 $\times$ 10$^{-4}$		\\
Ne               	& 18.18 $\times$ 10$^{-6}$   	& \multicolumn{5}{c}{}																			&						&						\\
He               	& 5.24 $\times$ 10$^{-6}$   	& \multicolumn{5}{c}{}																			&						&						\\
CH$_4$             & 1.70 $\times$ 10$^{-6}$   	& 190.6		& 4.6				& 35$^a$				& 714.29				& 1230.44					& 1.12 $\times$ 10$^{-5}$		& 1.73 $\times$ 10$^{-5}$		\\
Kr               	& 1.14 $\times$ 10$^{-6}$   	& 209.4		& 5.5 			& 30$^a$				& 400				& 637.52					& 4.20 $\times$ 10$^{-6}$		& 6.51 $\times$ 10$^{-6}$		\\
H$_2$               	& 0.55 $\times$ 10$^{-6}$   	& \multicolumn{5}{c}{}																			&						&						\\
CO              	& 1.25 $\times$ 10$^{-7}$   	& 133.1		& 3.5				& 37$^a$				& 1010.10				& 1794.95					& 1.16 $\times$ 10$^{-6}$		& 1.80 $\times$ 10$^{-6}$		\\
Xe 			& 0.87 $\times$ 10$^{-7}$ 	& 289.77		& 5.84			& 47$^c$				& 232.56				& 482.72					& 1.86 $\times$ 10$^{-7}$		& 2.89 $\times$ 10$^{-7}$		\\
\hline
\end{tabular}}}
\label{airgas}
\end{table*}

Based on this approach, we have elaborated a computational procedure (see Figure 1) aiming at calculating the volume of each gas dissolved in Lake Vostok (per unit of kg of lake water) and trapped in the newly forming air clathrates, as a function of the temporal evolution of the lake (in units of RT). The procedure is depicted as follows:

\begin{figure}
\label{diag}
\center
\resizebox{\hsize}{!}{\includegraphics[angle=0]{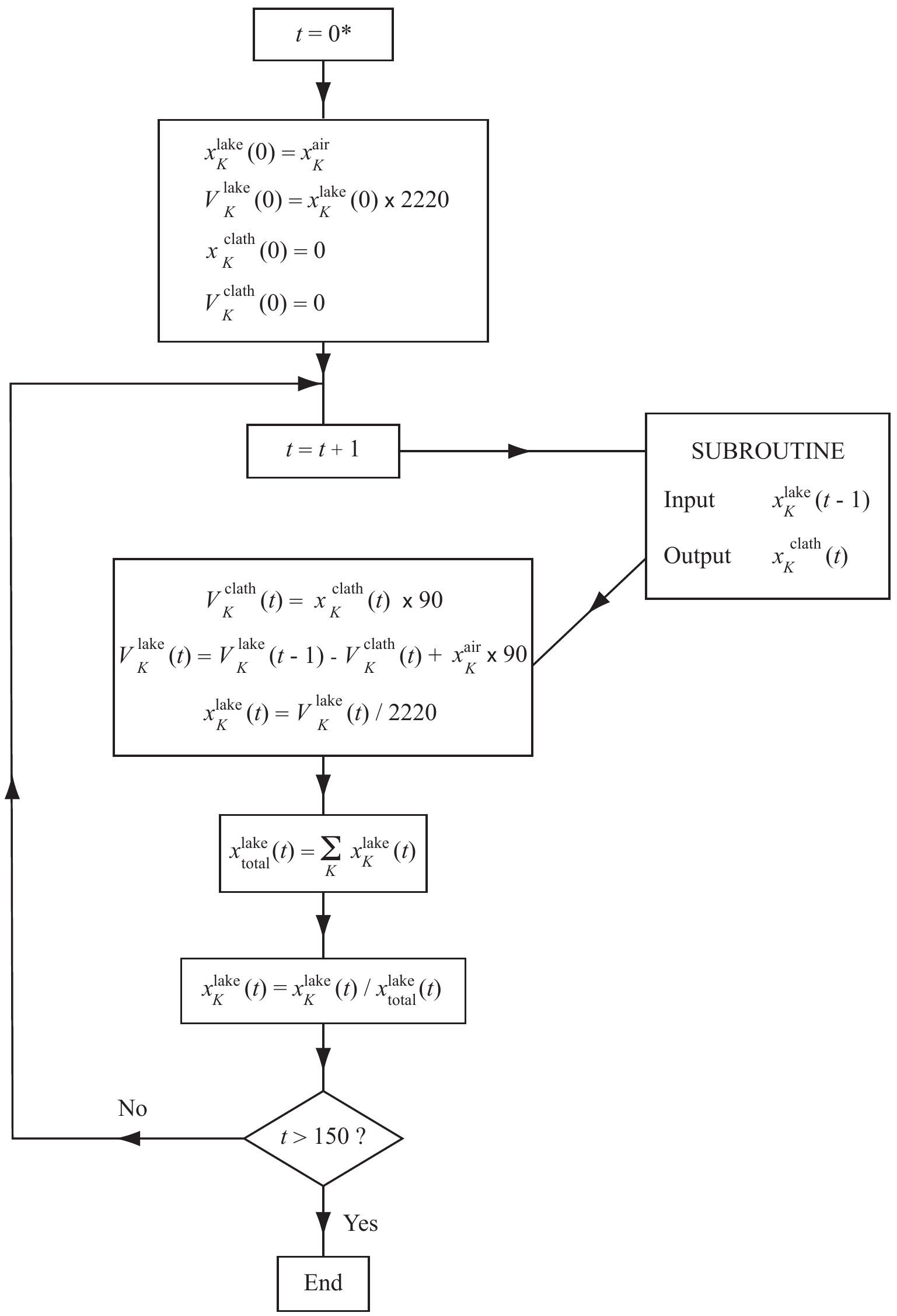}}
\caption{Structural scheme of our computational approach.}
\end{figure}

\begin{enumerate}

\item At time \textit t = 0, the mole fraction $x_{K}^{lake}(0)$ of species $K$, relative to the sum of all mole fractions of gases dissolved in the lake water, is given in Table \ref{airgas}. The corresponding volume of gas $V_{K}^{lake}(0)$ per unit of kg of water is determined. The mole fraction $x_{K}^{clath}(0)$ and the volume $V_{K}^{clath}(0)$ of the enclathrated species $K$ are set to zero;

\item  at \textit t = \textit t + 1, a subroutine (depicted in Section \ref{thermo}) allows us to compute the mole fraction $x_{K}^{clath}(t)$ of each enclathrated guest $K$ by using as input its volume (per unit of kg of water) in Lake Vostok at time \textit t - 1;

\item the new volumes $V_{K}^{clath}(t)$ and $V_{K}^{lake}(t)$ are calculated for each species $K$. The new mole fraction $x_{K}^{lake}(t)$ is obtained and normalized before being reintroduced in the next loop, until time \textit t reaches 150 RT, a value considered as infinity time. In the following, we consider that infinity time is the time at which the lake composition reaches steady state.

\end{enumerate}

\subsection{The statistical thermodynamic model}
\label{thermo}

To calculate the relative abundances of guest species incorporated in a clathrate from a coexisting gas of specified composition at given temperature and pressure, our subroutine follows the method described by Lunine \& Stevenson (1985), Thomas et al. (2007, 2008, 2009) and Mousis et al. (2010), which uses classical statistical mechanics to relate the macroscopic thermodynamic properties of clathrates to the molecular structure and interaction energies. It is based on the original ideas of van der Waals \& Platteeuw (1959) for clathrate formation, which assume that trapping of guest molecules into cages corresponds to the three-dimensional generalization of ideal localized adsorption. This approach is based on four key assumptions:

\begin{enumerate}
\item The host molecules contribution to the free energy is independent of the clathrate occupancy. This assumption implies that the guest species do not distort the cages;

\item the cages are singly occupied and guest molecules rotate freely within the cage;

\item guest molecules do not interact with each other;

\item classical statistics is valid, i.e., quantum effects are negligible.

\end{enumerate}

In this formalism, the fractional occupancy of a guest molecule $K$ for a given type $q$ ($q$~=~small or large) of cage in a clathrate structure can be written as

\begin{equation}
y_{K,q}=\frac{C_{K,q}~f_{hydro,K}}{1+\sum_{J}C_{J,q}f_{hydro,J}},
\label{eq1}
\end{equation}

\noindent where the sum in the denominator includes all the species that are present in the gas dissolved in lake water. $C_{K,q}$ is the Langmuir constant of species $K$ in the cage of type $q$, and $f_{hydro,K}$ the fugacity of guest species $K$ computed at the considered hydrostatic pressure (see Sec. \ref{fug}).

The Langmuir constant $C_{K,q}$ depends on the strength of the interaction between each guest species and each type of cage, and can be determined by integrating the molecular potential energy within the cavity as

\begin{equation}
C_{K,q}=\frac{4\pi}{k_B T}\int_{0}^{R_c}\exp\Big(-\frac{w_{K,q}(r)}{k_B T}\Big)r^2dr,
\label{eq2}
\end{equation}

\noindent where $R_c$ represents the radius of the cavity assumed to be spherical, $k_B$ the Boltzmann constant and $w_{K,q}(r)$ is the spherically averaged potential (here Kihara or Lennard-Jones potential) representing the interactions between the guest molecules $K$ and the H$_2$O molecules forming the surrounding cage $q$. This potential $w(r)$ can be written for a spherical guest molecule, as (McKoy \& Sinano\u{g}lu 1963)

\begin{equation}
w(r) = 2z\epsilon\Big[\frac{\sigma^{12}}{R_c^{11}r}\Big(\delta^{10}(r)+\frac{a}{R_c}\delta^{11}(r)\Big) - \frac{\sigma^6}{R_c^5r}\Big(\delta^4(r)+\frac{a}{R_c}\delta^5(r)\Big)\Big],
\label{eq3}
\end{equation}

\noindent with the mathematical function $\delta^N(r)$ in the form

\begin{equation}
\delta^N(r)=\frac{1}{N}\Big[\Big(1-\frac{r}{R_c}-\frac{a}{R_c}\Big)^{-N}-\Big(1+\frac{r}{R_c}-\frac{a}{R_c}\Big)^{-N}\Big].
\label{eq4}
\end{equation}

\noindent In Eqs. (\ref{eq3}) and (\ref{eq4}), $z$ is the coordination number of the cell.

Parameters $z$ and  $R_c$ depend on the type of the cage (small or large) and also on the structure of the clathrate. Indeed,
the air clathrates considered here may have structure I or II that are characterized by different arrangement of small and large cages of different sizes (Sloan \& Koh, 2008). The corresponding parameters $z$ and  $R_c$ are given in Table \ref{cageot}. The intermolecular parameters $a$, $\sigma$ and $\epsilon$ describing the guest molecule-water ($K-W$) interactions in the form of a Kihara or Lennard-Jones potential are listed in Table \ref{Kihara}. 

\begin{table}
\centering  \caption{Parameters for the clathrate cavities. $R_c$ is the radius of the cavity (values taken from Parrish \& Prausnitz 1972). $b$ represents the number of small ($b_s$) or large ($b_\ell$) cages per unit cell for a given structure of clathrate (I or II) with volume $V_c$, $z$ is the coordination number in a cavity.}
\begin{tabular}{lcccc}
\hline \hline
Clathrate structure 		& \multicolumn{2}{c}{I} 			& \multicolumn{2}{c}{II} 			\\
\hline
$N_W$				& \multicolumn{2}{c} {46}	 		& \multicolumn{2}{c} {136} 		\\
$V_c$ (\AA$^3$) 		& \multicolumn{2}{c}{$12.0^3$} 	& \multicolumn{2}{c}{$17.3^3$} 	\\
Cavity type     			& small     		& large     			& small     		& large 			\\
$R_c$ (\AA)     			& 3.975     	& 4.300     		& 3.910     	& 4.730 			\\
$b$             			& 2         		& 6         			& 16       		& 8     			\\
$z$             			& 20        		& 24        			& 20        		& 28    			\\
\hline
\end{tabular}
\label{cageot}
\end{table}

\begin{table*}
\centering \caption{Parameters for Kihara and Lennard-Jones potentials. $\sigma_{K-W}$ is the Lennard-Jones diameter, $\epsilon_{K-W}$ is the depth of the potential well, and $a_{K-W}$ is
the radius of the impenetrable core, for the guest-water pairs.}
\begin{tabular}{lcccc}
\hline \hline
Molecule   & $\sigma_{K-W}$ (\AA)& $\epsilon~_{K-W}/k_B$ (K)& $a_{K-W}$ (\AA) 	& Reference		\\
\hline
CO$_2$			& 2.97638		& 175.405		& 0.6805	&	Sloan \& Koh (2008)				\\
N$_2$			& 3.13512		& 127.426		& 0.3526	&	Sloan \& Koh (2008)				\\
O$_2$	     		& 2.7673     	& 166.37     	& 0.3600 	&	Parrish \& Prausnitz (1972)		\\
CH$_4$	     		& 3.14393     	& 155.593     	& 0.3834 	&	Sloan \& Koh (2008)				\\
CO			       	& 3.1515     	& 133.61     	& 0.3976 	&	Mohammadi et al. (2005)			\\
Ar              			& 2.9434     	& 170.50     	& 0.184 	&	Parrish \& Prausnitz (1972)		\\
Kr             			& 2.9739     	& 198.34     	& 0.230 	&	Parrish \& Prausnitz (1972)		\\
Xe              		& 3.32968     	& 193.708     	& 0.2357 	&	Sloan \& Koh (2008)				\\
\hline
\end{tabular}
\label{Kihara}
\end{table*}

Finally, the mole fraction $x^{clat}_K$ of a guest molecule $K$ in a clathrate of a given structure (I or II) can be calculated with respect to the whole set of species considered in the system as

\begin{equation}
x^{clat}_K=\frac{b_s y_{K,s}+b_\ell y_{K,\ell}}{b_s \sum_J{y_{J,s}}+b_\ell \sum_J{y_{J,\ell}}},
\label{eq5}
\end{equation}

\noindent where $b_s$ and $b_l$ are the number of small and large cages per unit cell, respectively, for the clathrate structure under consideration. Note that the sum of the mole fractions of the enclathrated species is normalized to 1. We also neglect Ne, He and H$_2$ in our computation of the composition of clathrates as these species have low trapping propensities in these structures (Lunine \& Stevenson 1985; Sloan \& Koh 2008).

\subsection{Determination of the fugacities}
\label{fug}

To determine $f_{hydro,K}$, we first calculate the specific volume $\nu$ of the considered mixture via the Redlich-Kwong equation of state (Redlich \& Kwong 1949):
 
\begin{equation}
P =\frac{R~T}{\nu - b}-\frac{a}{\sqrt T~\nu~(\nu + b)},
\label{eq1a}
\end{equation}

\noindent with

\begin{equation}
a = 0.42748~\frac{R^2~T_{c}^{2.5}}{P_{c}}~{\rm and}~b~=~0.08664~\frac{R~T_{c}}{P_{c}},
\label{eq1b}
\end{equation}

\noindent where $R$ is the gas constant, $T$ the ambient temperature, $T_{c}$ and $P_{c}$ the critical temperature and pressure of the substance (see Table \ref{airgas}), and $P$ its vapor pressure. Just as any other cubic equation of state, the Redlich-Kwong EOS has to be applied only to pure substances. For mixtures, however, the same equation is applied but certain mixing rules are applied to obtain parameters $a$ and $b$, which are functions of the properties of the pure components. This corresponds to the creation of a new ``pseudo'' pure substance that has the average properties  of the mixture. Following Redlich \& Kwong (1949), we use the mixing rules:

\begin{equation}
a_m = \Sigma_i \Sigma_j~x_i^{lake}~x_j^{lake}~a_{ij}~~{\rm with}~~a_{ij}~=~\sqrt {a_i a_j}, \nonumber
\label{eq1b2}
\end{equation}
\begin{equation}
b_m = \Sigma_i~x_i^{lake}~b_i.
\label{eq1b2}
\end{equation}

Here the mixture pressure $P_m$ dissolved in water corresponds to the sum of the individual gas pressures $P_K$. Each $P_K$ is expressed as a function of the Henry's law coefficient as:

\begin{equation}
P_K = H_{hydro,K} \times y^{lake}_K,
\label{eq1c}
\end{equation}

\noindent where $H_{hydro,K}$ is the Henry's law constant calculated for species $K$ at the lake's hydrostatic pressure and ambient temperature, and $y^{lake}_K$ the mole fraction of species $K$ expressed per unit of volume of water. This term is derived from (Krichevsky \& Kasarnovsky 1935):

\begin{equation}
ln \left (\frac{H_{hydro,K}}{H_{1atm,K}} \right)~=~\frac{\overline{V}_{K}}{R~T} (P_{hydro} - 1.013 \times 10^5),
\label{eq1d}
\end{equation}

\noindent where $P_{hydro}$ is the lake's mean hydrostatic pressure ($\sim$35 MPa), $\overline{V}_{K}$ the partial molar volume of species $K$ and $H_{1atm,K}$ the Henry's constant determined at 1 atmosphere and at 0$\degre$C (see Table \ref{airgas}).

The fugacity of the mixture at 1 atm pressure $f_{1atm,m}$ is then determined from the following relation (Redlich \& Kwong 1949):

\begin{eqnarray}
ln~f_{1atm,m}= \frac{b_m}{\nu_m - b_m} + ln~\frac{R~T}{\nu_m - b_m} - \nonumber \\ 
\frac{a_m}{R~T^{3/2}} \left (\frac{1}{\nu_m + b_m}+\frac{1}{b_m}~ln~\frac{(\nu_m + b_m)}{\nu_m} \right).
\label{eq1e}
\end{eqnarray}

\noindent Because it is calculated at the hydrostatic pressure, $f_{hydro,m}$ is related to $f_{1atm,m}$ via (Miller 1974):

\begin{eqnarray}
ln~\frac{f_{hydro,m}}{f_{1atm,m}}=~-~n~ln~a_W \nonumber \\ 
-~\frac{P_{hydro} - 1.013 \times 10^5}{R~T} (n~V_{H_2O}~-~V_{c,m}),
\label{eq1f}
\end{eqnarray}

\noindent where $a_W$ is the activity of water relative to pure liquid water (here $a_W$ is $\sim$1), n is the moles of water per mole of species $K$ in the clathrate (n $\sim$6), $V_{H_2O}$ is the molar volume of liquid water (1.8 $\times$ 10$^{-5}$ m$^3$) and $V_{c,m}$ is the volume of clathrate that contains 1 mole of substance $m$ (here approximated to be dominated by N$_2$). Finally the fugacity coefficient $\phi$ of the mixture, which is defined as the ratio of $f_{hydro,m}$ to $P_{hydro,m}$, allows to retrieve the individual fugacities following

\begin{equation}
f_{hydro,K} = \phi \times P_K.
\label{eq1g}
\end{equation}

\section{Results}
\label{results}

\subsection{Evolution of clathrate composition}
\label{clat_comp}

The composition of structure I and II clathrates that might form from the air dissolved in Lake Vostok has been computed using the approach presented above. If the lake has experienced less than $\sim$25 RT, the different species remain completely in solution. At longer times, clathrates start to form. Figures 2 and 3 display the compositions of the clathrates of same structure successively formed at different epochs of the lake evolution. At each timestep, the total volume of enclathrated gases is 90 cm$^3$. Irrespective of the clathrate structure, the mole fractions of volatiles trapped in clathrates converge toward their atmospheric mole fractions at infinite RT as the result of mass conservation (MK03). Steady state is reached 40--80 RT after clathrate formation. In both structures, N$_2$ and O$_2$ remain the main gases trapped in clathrates with variations of their mole fractions depending on the considered structure and epoch of clathrate formation. In the case of structure I clathrates, the mole fractions of trapped N$_2$ and O$_2$ change slightly over time. Indeed, the mixing ratio of N$_2$ decreases from $\sim$0.86 in clathrates formed at early epochs to 0.78 in clathrates formed at infinite time whereas that of O$_2$ simultaneously increases from $\sim$0.12 to 0.21 in clathrates successively formed during the same time interval. In the case of Structure II clathrates, the mole fractions of N$_2$ and O$_2$ incorporated in clathrates are even less sensitive to their formation epoch in the lake. Thus, the mole fraction of N$_2$ increases from 0.74 to 0.78 while that of O$_2$ decreases from 0.24 to 0.21 in clathrates formed between first formation and steady state.

\begin{figure}
\label{clat1}
\center
\resizebox{\hsize}{!}{\includegraphics[angle=0]{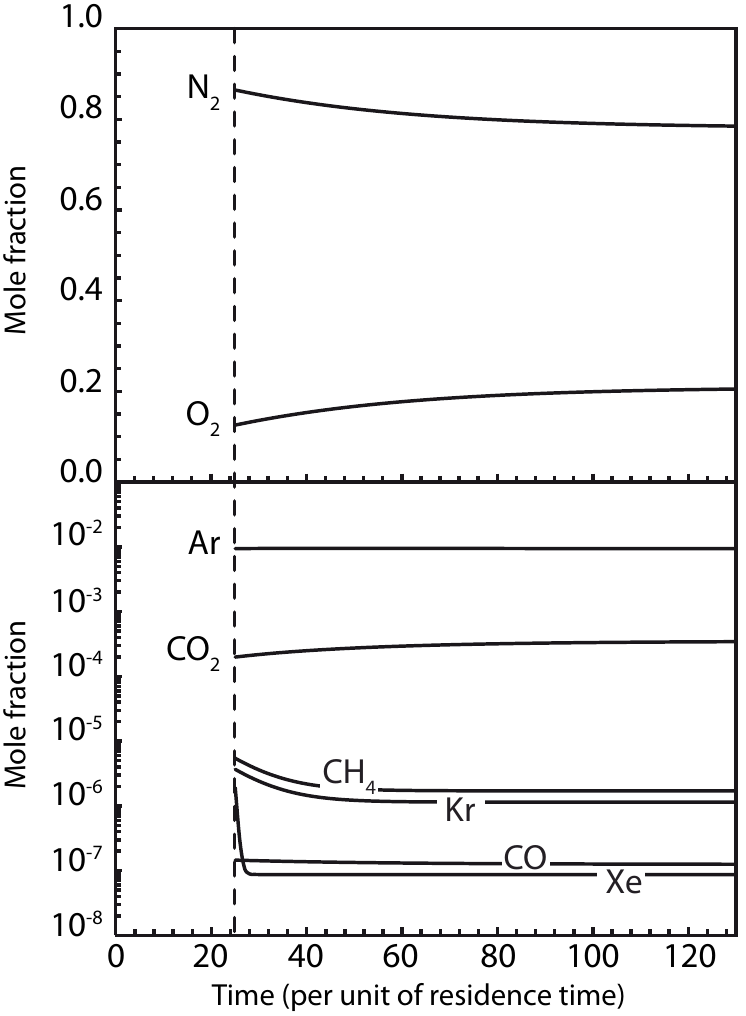}}
\caption{Mole fractions of N$_2$, O$_2$, Ar, CO$_2$, CH$_4$, Kr, Xe  and CO trapped in structure I clathrates formed at different ages of the lake, with age expressed in units of RT. Clathrate formation starts at {$\sim$25 RT} (see text).}
\end{figure}

\begin{figure}
\label{clat2}
\center
\resizebox{\hsize}{!}{\includegraphics[angle=0]{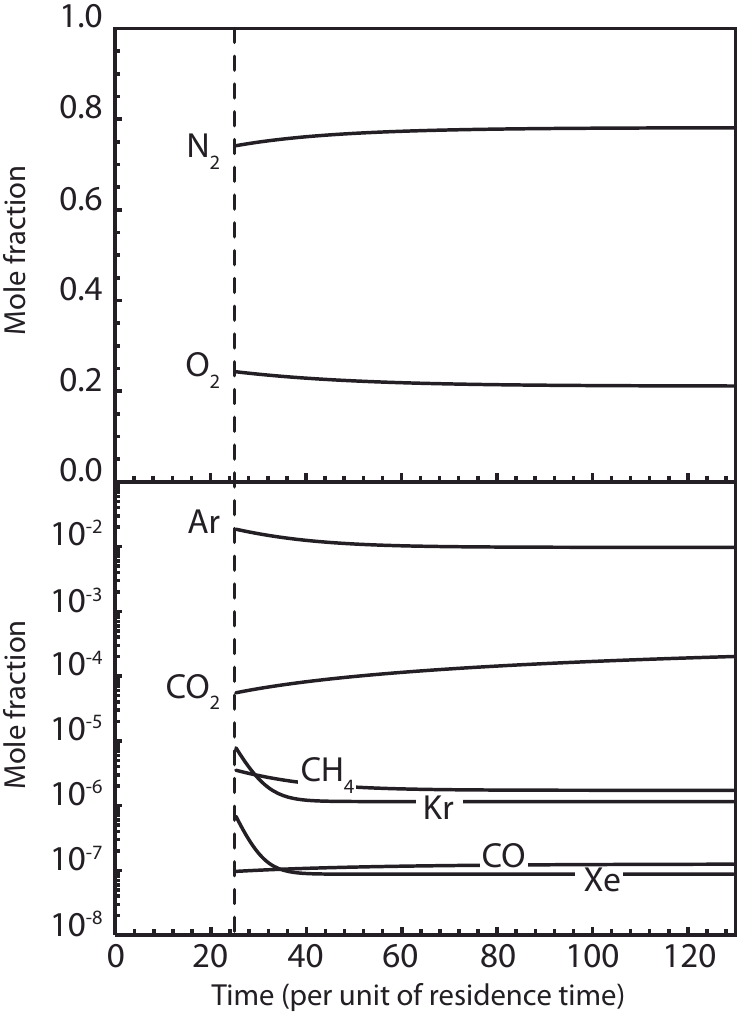}}
\caption{Mole fractions of N$_2$, O$_2$, Ar, CO$_2$, CH$_4$, Kr, Xe  and CO trapped in structure II clathrates formed at different ages of the lake, with age expressed in units of RT. Clathrate formation starts at $\sim$25 RT (see text).}
\end{figure}

Moreover, we note that, with an enrichment factor of $\sim$22 compared to the atmospheric abundance (see Table \ref{airgas}), Xe is  efficiently sequestrated at early formation epochs ($\sim$25 RT) in structure I clathrates. To a lesser extent, a similar trend is calculated for Kr and CH$_4$ whose mole fractions are both enriched by a factor of $\sim$3.2 at early epochs of structure I clathrate formation in the lake. The same remarks apply for the trapping of volatiles in structure II clathrates, for which the mole fractions of Xe, Kr and CH$_4$ are enriched by factors $\sim$78, 6.7 and 2 at early formation epochs. On the other hand, the trapping efficiencies of Ar and CO in clathrate depend on the considered structure. In structure I, the mole fraction of Ar remains similar to its atmospheric value, irrespective of the considered RT. In structure II, however, we note that Ar is initially enriched by a factor of $\sim$2 in clathrate. In addition, CO is initially slightly enriched by a factor of $\sim$1.2 in structure I clathrate while it is impoverished by a factor of $\sim$0.8 in structure II clathrate. In both structures and at early epochs of clathrate formation, the mole fraction of trapped CO$_2$ is lower than its atmospheric value (factors of $\sim$0.6 and 0.15 in structures I and II, respectively). In both cases, the mole fractions of all volatiles trapped in clathrate progressively converge with time towards their atmospheric values.

\subsection{Evolution of the lake composition}
\label{lake_comp}

Figures 4--7 illustrate the evolution of the composition of the gas phase dissolved in Lake Vostok as a function of its RT in the cases of structures I and II clathrates. In both cases, the mole fractions of the different volatiles relative to the sum of all mole fractions of dissolved gases and present in the lake linearly increase as long as its lifetime has not exceeded $\sim$25 RT. After this epoch, these volatiles have reached their solubility limit in the lake and clathrate forms from the gas supplied in excess by ice melting. The resulting volatile content of the lake at a given time is then controlled by the balance between the sequestration of gas by the forming clathrate and the air supplied by the melting ice.

\begin{figure}
\label{lac1}
\center
\resizebox{\hsize}{!}{\includegraphics[angle=0]{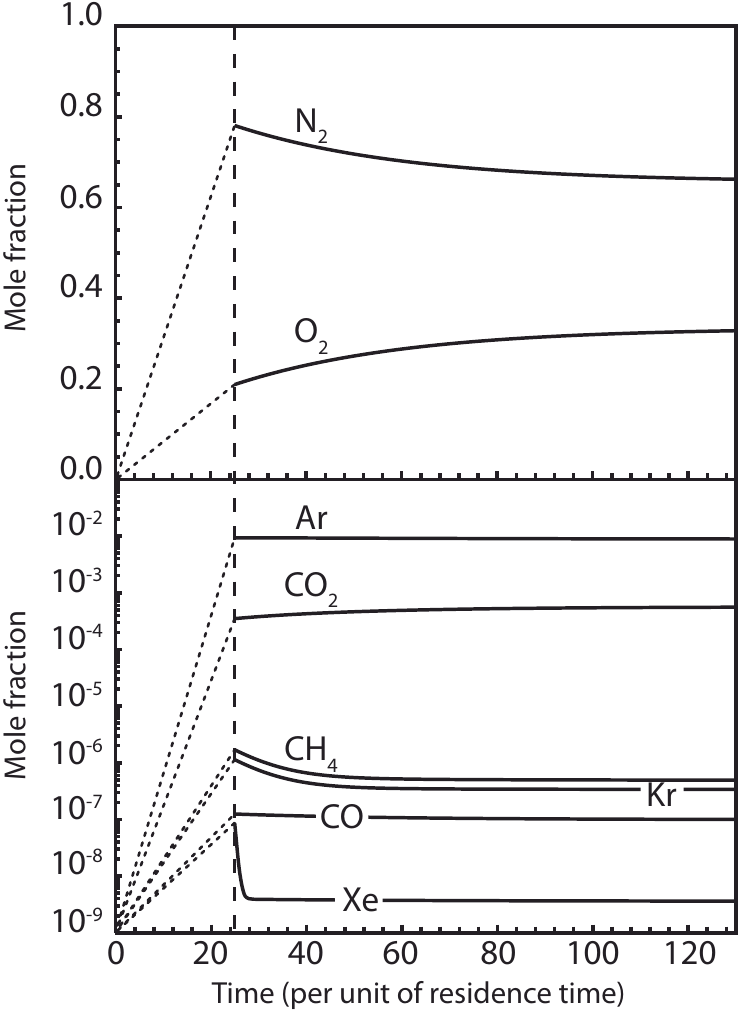}}
\caption{Mole fractions of N$_2$, O$_2$, Ar, CO$_2$, CH$_4$, Kr, CO and Xe dissolved in Lake Vostok calculated in the case of structure I clathrate formation and expressed as a function of the age of lake, with age expressed in units of RT.}
\end{figure}

\begin{figure}
\label{lac2}
\center
\resizebox{\hsize}{!}{\includegraphics[angle=0]{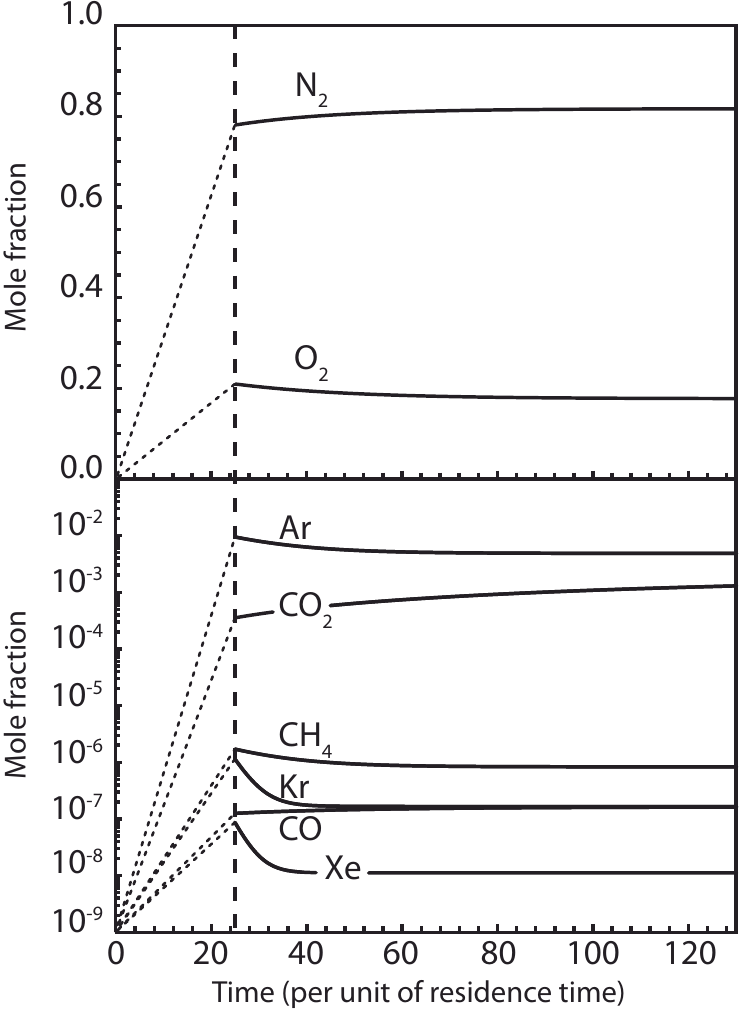}}
\caption{Mole fractions of N$_2$, O$_2$, Ar, CO$_2$, CH$_4$, Kr, CO and Xe dissolved in Lake Vostok calculated in the case of structure II clathrate formation and expressed as a function of the age of lake, with age expressed in units of RT.}
\end{figure}

\begin{figure}
\label{lac1a}
\center
\resizebox{\hsize}{!}{\includegraphics[angle=0]{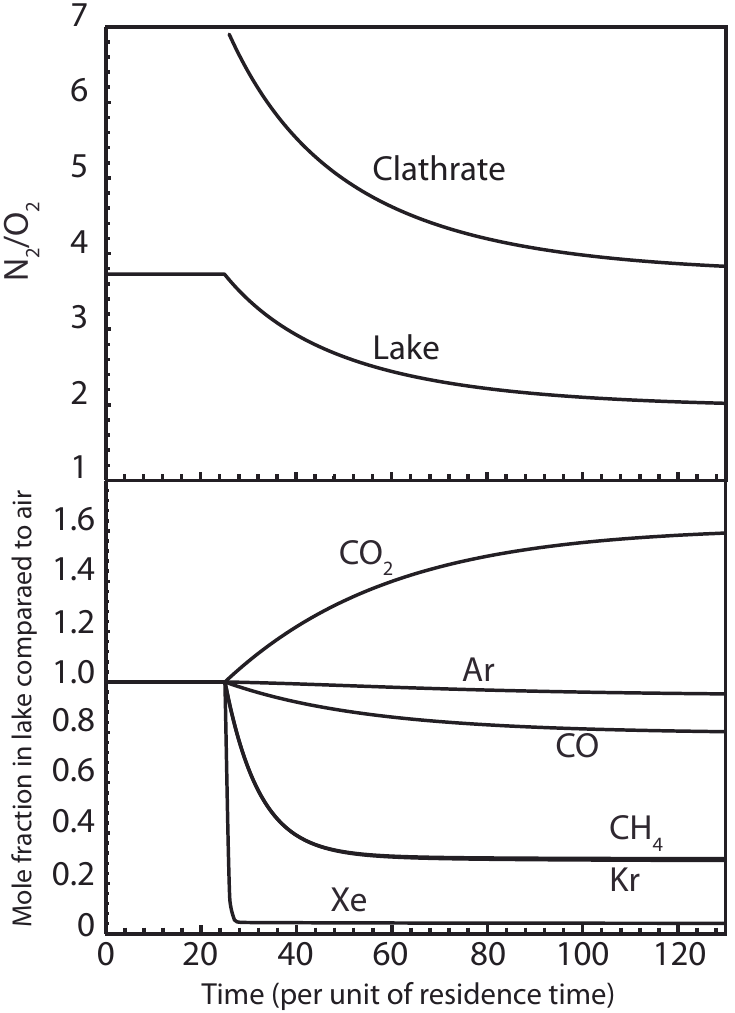}}
\caption{Top panel:  evolution of the N$_2$/O$_2$ ratio in lake and in structure I clathrate as a function of time. Bottom panel: evolution of the mole fractions of Ar, CO$_2$, CH$_4$, Kr, CO and Xe relative to their atmospheric values in lake as a function of time in case of structure I clathrate formation.}
\end{figure}

\begin{figure}
\label{lac2a}
\center
\resizebox{\hsize}{!}{\includegraphics[angle=0]{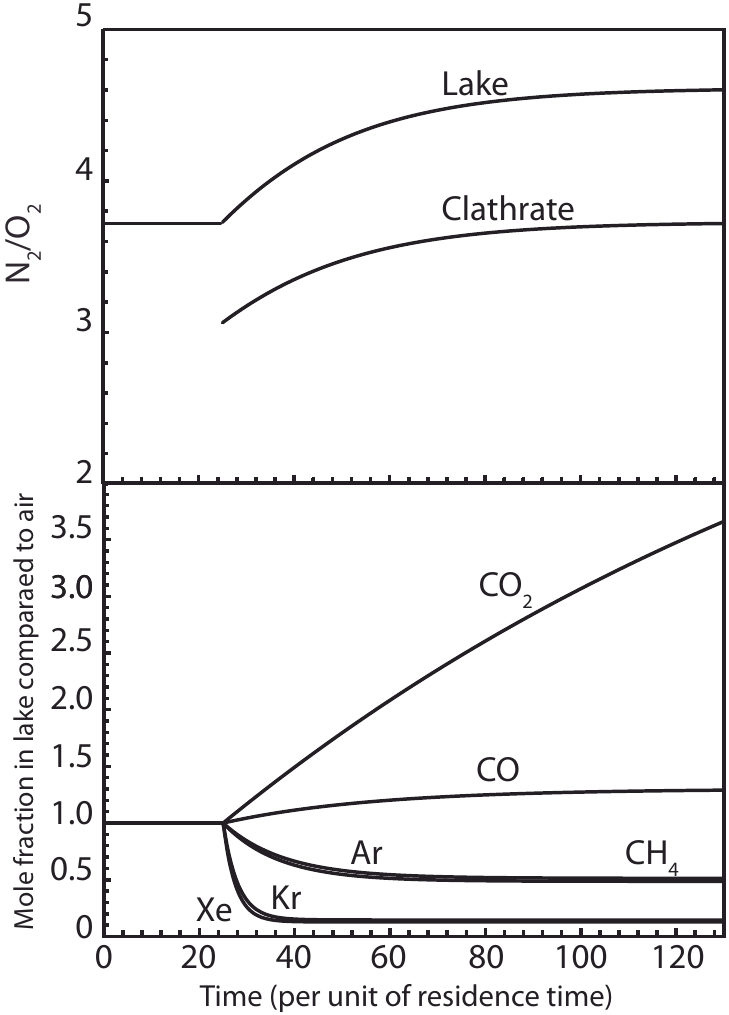}}
\caption{Top panel:  evolution of the N$_2$/O$_2$ ratio in lake and in structure II clathrate as a function of time. Bottom panel: evolution of the mole fractions of Ar, CO$_2$, CH$_4$, Kr, CO and Xe relative to their atmospheric values in lake as a function of time in case of structure II clathrate formation. }
\end{figure}

In the two cases, N$_2$ and O$_2$ remain the main gases dissolved in water. When considering the formation of structure I clathrate, and irrespective of the considered epoch, the N$_2$ mole fraction in the lake water decreases from its atmospheric value (0.78) down to 0.66 at steady state. On the other hand, the O$_2$ mole fraction slightly increases with time in water from 0.21 (the atmospheric mole fraction) to 0.33 in the lake. As a result, the N$_2$/O$_2$ ratio in the lake water is much lower than the N$_2$/O$_2$ ratio in structure I clathrate (see Fig. 6). Because the mole fractions of Xe, Kr, CH$_4$ and CO are all enriched in structure I clathrate, they become correspondingly depleted by factors $\sim$0.04, 0.3, 0.3 and 0.8 in the lake water compared to their atmospheric abundances (see Table \ref{airgas}), and irrespective of the considered RT. In contrast, the mole fraction of Ar in the lake water remains very close (by a factor of $\sim$0.95) to its atmospheric value while that of CO$_2$ rapidly increases with time up to 1.6 times its atmospheric mole fraction at 200 RT.

In the case of structure II clathrate formation in Lake Vostok, the N$_2$ mole fraction in the lake water increases from its atmospheric value (0.78) up to 0.82 at steady state. Inversely, the O$_2$ mole fraction decreases in water from its atmospheric value (0.21) down to 0.18 at steady state. The resulting N$_2$/O$_2$ ratio is then significantly higher in the lake water than the N$_2$/O$_2$ ratio in structure II clathrate, irrespective of the considered RT (see Fig. 7). The mole fractions of Xe, Kr, CH$_4$, and Ar also become rapidly impoverished by the factors $\sim$0.1, 0.15, 0.5 and 0.5 in the lake water, respectively, compared to the atmospheric values (see Table \ref{airgas}), and irrespective of the considered RT. On the other hand, CO becomes moderately enriched by a factor $\sim$1.3 in the water compared to its atmospheric value. Moreover, due to its lower propensity to be trapped in structure II clathrate, the CO$_2$ mole fraction in the lake water presents the highest enrichment (up to a factor of $\sim$5 at 200 RT) compared to its atmospheric abundance.
     
\subsection{Clathrate density}
\label{density}

Once formed, clathrate may either float on the surface or sink to the lake floor according to its density with respect to that of the lake water. To investigate this point, we have computed the densities of structures I and II clathrates formed in Lake Vostok by following the method depicted in Sloan \& Koh (2008; Eq. 5.2.1 p268). We find that, in both cases, these clathrates are all lighter than liquid water (see Fig. 8). This conclusion is in agreement with the results obtained by MK03, although we do not calculate the same clathrate composition. However, air clathrates have not been observed in the accreted ice above the lake (Siegert et al. 2000). To explain this, MK03 argued that significant amounts of CO$_2$ could have been produced in Lake Vostok and subsequently incorporated in clathrates, thus changing their density with respect to that of clathrates simply formed in contact with air entering to the lake through the ice sheet. These clathrates would have sunk to the lake floor as a result of their higher density. We computed the minimum fraction of CO$_2$ present in the lake necessary to be trapped in air clathrates so that the clathrates sink to the lake floor. In the case of structure I clathrate, our model suggests that the minimum mole fraction of CO$_2$ present in Lake Vostok must be around $\sim$0.05, with a corresponding mole fraction that is similar in clathrate (see Fig. 8). In the case of structure II clathrate, the minimum mole fraction of CO$_2$ needed to induce a higher clathrate density than that of lake water largely exceeds 0.5, implying that this species becomes dominant in the lake. With such a high mole fraction of CO$_2$ in the lake, structure I clathrate should be the most stable form (Sloan \& Koh 2008) so we should expect a transition from structure II to structure I clathrate, which would present an even higher density.

\begin{figure}
\label{dens}
\center
\resizebox{\hsize}{!}{\includegraphics[angle=0]{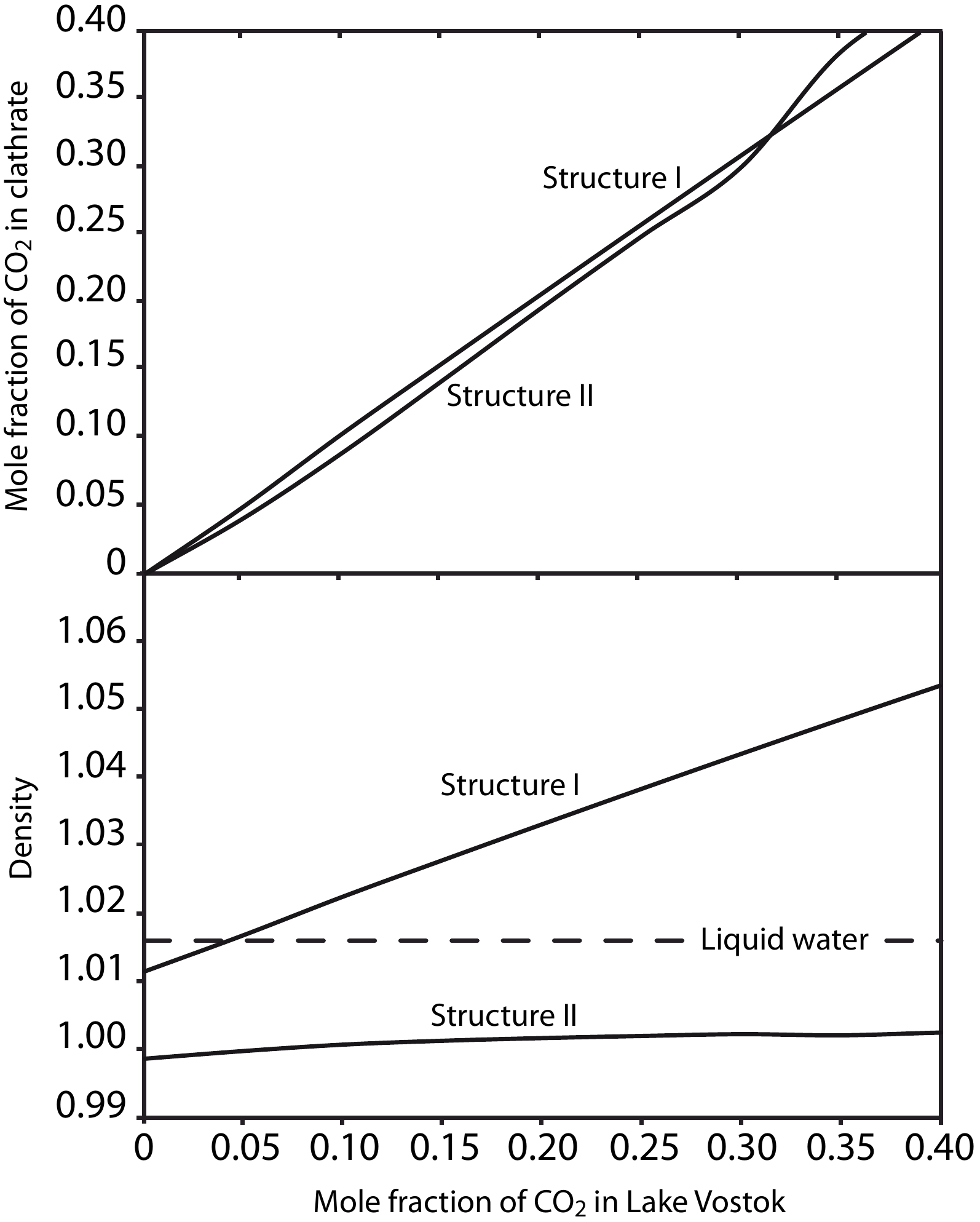}}
\caption{Top panel: mole fraction of CO$_2$ incorporated in structures I and II clathrates as a function of the mole fraction of CO$_2$ dissolved in Lake Vostok. Bottom panel: densities of structures I and II clathrates as a function of the mole fraction of CO$_2$ dissolved in Lake Vostok. The horizontal dashed line corresponds to the density of liquid water in Lake Vostok (Lipenkov \& Istomin 2001).}
\end{figure}

\section{Discussion}
\label{discuss}

The composition of Lake Vostok calculated with the present approach displays significant differences with the one predicted by MK03. These differences come from both the consideration of a larger set of molecules in the gas phase dissolved in the lake (we include CH$_4$ and CO) and from the use of a more sophisticated model to compute the mole fractions of all species simultaneously trapped in the clathrates, in a self consistent way. As a consequence, our calculations suggest enhancements of the Xe and Kr mole fractions trapped in clathrate by factors of $\sim$ 7.6--26.9 and 2.0--4.2, respectively, compared to the values previously determined by MK03. Our model therefore suggests that Xe and Kr abundances are $\sim$0.04--0.13 and 0.24--0.5 times those predicted by MK03 in Lake Vostok water. Our model also predicts an impoverishment of the Ar abundance in the lake water (factor $\sim$0.5--0.95 compared to the atmospheric abundance), in contrast with the enrichment (factor $\sim$1.4) predicted by MK03. Moreover, our calculations show that CO$_2$ is significantly enriched in the lake water (factor $\sim$1.6--5 at 200 RT following the considered clathrate structure) compared to their atmospheric abundances. In both cases, CH$_4$ appears depleted in the lake water compared to its atmospheric mole fraction (factor $\sim$0.3--0.5). Depending on the considered structure, CO can be found in the lake water moderately depleted (factor $\sim$0.8 in the case of structure I clathrate) or enriched (factor $\sim$1.3 in the case of structure II clathrate) compared to its atmospheric value. It is important to note that, in absence of large amounts of CO$_2$ dissolved in the liquid water, since the composition of the dissolved gas is dominated by N$_2$ and because N$_2$ forms structure II clathrate (Sloan \& Koh 2008), we expect that structure II clathrate is the most likely to be formed in Lake Vostok.

Note that the CO$_2$ atmospheric abundance considered in this work corresponds to the one measured nowadays in the Earth's atmosphere, which is affected by industrial emissions. In the same way, the CO$_2$ atmospheric abundance is estimated to have been 30$\%$ lower than the current value some 300 years ago and might have even been 10$\%$ higher 1--15 million years ago. Moreover, it has been inferred that extreme fractionation of gases could occur by formation of clathrates in Vostok ice due, at least partly, to different diffusion coefficients in ice of the considered gases (Ikeda et al. 1999). As a consequence, using the present atmospheric abundances for determining the composition of clathrates formed from air trapped for perhaps thousands of year in the ice sheet above the lake could be questionable. However, in situ measurements have shown that the N$_2$/O$_2$ ratio delivered to the lake at the bottom of the ice sheet is similar to the present atmospheric ratio (Ikeda et al. 1999). Similarly, for CO$_2$, it has been shown that the fractionation process in Vostok ice is smoothed out with depth, regaining the initial mean atmospheric concentration below the clathrate zone formation (Luthi et al. 2010). This information is unfortunately lacking for the other gases. However, additional tests performed by changing the CO$_2$ composition in our approach show that, in any case, these variations do not affect our conclusions concerning the efficiency of CO$_2$ trapping in clathrates and its consecutive impoverishment in Lake Vostok.

The comparison of our calculated CO$_2$ and CH$_4$ mole fractions in Lake Vostok with future in situ measurements will allow to discriminate between their different supply sources. Indeed CH$_4$ and CO$_2$ abundances in Lake Vostok will vary according to the amount and type of biological and abiotic activity affecting the speciation of carbon. If biological activity and if water-rock interactions are minimal, then the concentration of both of these gases should approach the amount present in the gases at the time of Lake Vostok's formation. However, microbes, if present (as is highly likely -- see Priscu et al. 1999; Karl et al. 1999), would significantly alter the abundances of these gases in the fluids. If water-rock interactions are minimal, then microbes should likely consume methane and other reducing compounds, oxidizing them to form CO$_2$.  As such, the CO$_2$ content of the fluid should increase relative to the amount predicted, and the CH$_4$ content should decrease, especially because O$_2$ would be abundant in these oxidizing waters.  However, if water-rock interactions are present within the lake, then chemolithotrophic microbes might survive in these environments, as suggested by Bulat et al. (2004), and may perhaps accelerate oxidation of rock material with dissolved O$_2$.  The ability of microbes to do so would be contingent on the composition of the rocks at the base of the lake.

Lake Vostok may be compared favorably with Europa in terms of pressure, temperature, and potentially water composition. The pressure at the water--ice layer of Lake Vostok would be comparable to a water-ice layer on Europa occurring at 30 km. Furthermore, the presently calculated quantity of O$_2$ dissolved within Lake Vostok is also similar to those predicted by Hand et al. (2006) and Greenberg (2010) on Europa, in which O$_2$ would be an abundant constituent of the water. Although the composition of gases present on Europa is unclear, O$_2$ has been detected on the surface of Europa (Hall et al. 1995; Carlson et al. 1999). This O$_2$ is likely transported from the surface to the subsurface ocean.  The gas composition of other species is less certain, however it is also highly likely that the original icy material from which Europa formed also had several gas clathrates, including those of the noble gases (Kereszturia \& Keszthelyib 2012; Mousis \& Gautier 2004; Mousis \& Alibert 2006; Mousis et al. 2011).  These gases would be unlikely to have escaped from Europa, and hence should still be present in some quantity in the subsurface ocean. As the water-rock interactions at the base of the lake are unlikely to produce abundant hydrothermal systems as the underlying rock is felsic or silicic in composition (Van der Fliert et al. 2008), there is little likelihood of significant acidification occurring as O$_2$ reacts with H$_2$S produced by hydrothermal systems, as suggested for Europa by Pasek \& Greenberg (2012). For these reasons, water in Lake Vostok and in Europa is likely close to neutral in pH and highly oxidizing. Because of the similitude between Lake Vostok and Europa, our approach would thus be a very useful tool for accurate predictions of the composition of Europa's internal ocean, if validated by comparison with in situ measurements in Lake Vostok.

\section*{Acknowledgements}

We are grateful to Christopher McKay and an anonymous Referee for the careful reading and the valuable comments and suggestions to improve this paper.


\begin{thebibliography}{00}

\bibitem[Anderson (2002)]{2002Nat} Anderson, G. K. (2002)\ Solubility of Carbon Dioxide in Water under Incipient Clathrate Formation Conditions. \emph{J. Chem. Eng. Data} 47: 219-222.

\bibitem[Bell et al.(2002)]{2002Natur.416..307B} Bell, R.~E., Studinger, M., Tikku, A.~A., Clarke, G.~K.~C., Gutner, M.~M., Meertens, C.\ (2002)\ Origin and fate of Lake Vostok water frozen to the base of the East Antarctic ice sheet.\ \emph{Nature} 416:307-310.

\bibitem[Bulat et al.(2004)]{2004IJAsB...3....1B} Bulat, S.~A., and 10 colleagues (2004)\ DNA signature of thermophilic bacteria from the aged accretion ice of Lake Vostok, Antarctica: implications for searching for life in extreme icy environments.\ \emph{International Journal of Astrobiology}  3:1-12.

\bibitem[Carlson et al.(1999)]{1999Sci...283.2062C} Carlson, R.~W., and 13 colleagues (1999)\ Hydrogen Peroxide on the Surface of Europa.\ \emph{Science} 283: 2062.

\bibitem[Greenberg(2010)]{2010AsBio..10..275G} Greenberg, R.\ (2010)\ Transport Rates of Radiolytic Substances into Europa's Ocean: Implications for the Potential Origin and Maintenance of Life.\ \emph{Astrobiology} 10:275-283.

\bibitem[Hall et al.(1995)]{1995Natur.373..677H} Hall, D.~T., Strobel, D.~F., Feldman, P.~D., McGrath, M.~A., Weaver, H.~A. (1995)\ Detection of an oxygen atmosphere on Jupiter's moon Europa.\ \emph{Nature} 373: 677-679.

\bibitem[Hand et al.(2006)]{2006AsBio...6..463H} Hand, K.~P., Chyba, C.~F., Carlson, R.~W., Cooper, J.~F.\ (2006)\ Clathrate Hydrates of Oxidants in the Ice Shell of Europa.\ \emph{Astrobiology} 6:463-482.

\bibitem[Ikeda et al. (1999)]{2006AsBio...6..3H} Ikeda, T., Fukazawa, H., Mae, S., Pepin, L., Duval, P., Champagnon, B., Lipenkov, V.Y., Hondoh, T. (1999) Extreme fractionation of gases caused by formation of clathrate hydrates in Vostok Antarctic ice. \emph{Geophysical Research Letters}, 26:91-94.

\bibitem[Jouzel et al.(1999)]{1999Sci...317..793J} Jouzel, J., Petit, J. R., Souchez, R., Barkov, N. I., Lipenkov, V. Ya, Raynaud, D., Stievenard, M., Vassiliev, N. I., Verbeke, V, Vimeux, F. (1999) More than 200 meters of lake ice above subglacial Lake Vostok, Antarctica. \emph{Science} 286:2138-2141.

\bibitem[Kapitsa et al.(1996)]{1996Natur.381..684K} Kapitsa, A.~P., Ridley, J.~K., de Q.~Robin, G., Siegert, M.~J., Zotikov, I.~A.\ (1996)\ A large deep freshwater lake beneath the ice of central East Antarctica.\ \emph{Nature} 381:684-686.

\bibitem[Kapit al.(1996)]{1996.381..684K} Karl, D. M., Bird, D. F., Bjorkman, K., Houlihan, T., Schakelford, R., Tupas, L. (1999) Microorganisms in the accreted ice of Lake Vostok, Antarctica. \emph{Science} 286:2144-2147.

\bibitem[Ka al.(1996)]{199381..684K} Krichevsky, I. R., Kasarnovsky, J. S. (1935) Thermodynamical calculations of solubilities of nitrogen and hydrogen in water at high pressures. \emph{J. Am. Chem. Soc.} 57: 2168-2171.

\bibitem[Kennan (1990)]{1991..684K} Kennan, R. P., Pollack, G. L. (1990) Pressure dependence of the solubility of nitrogen, argon, krypton, and xenon in water. \emph{J. Chem. Phys.} 93(4): 2724.

\bibitem[Kereszturi and Rivera-(2012)]{2012Ica.221..289K} Kereszturi, A., Keszthelyib, Z. (2012). Astrobiological implications of chaos terrains on Europa to help targeting future missions. \emph{Planetary and Space Science}, in press.

\bibitem[Lide(2002)]{2002crc..book.....L} Lide, D.~R.\ (2002)\ CRC Handbook of chemistry and physics : a ready-reference book of chemical and physical data. 83rd ed., by David R.~Lide.~\emph{Boca Raton: CRC Press} ISBN 0849304830.

\bibitem[Lipenkov (2001)]{2001Natur.381..684K} Lipenkov, V. Ya., Istomin, V. A. (2001) On the stability of air clathrate-hydrate crystals in subglacial Lake Vostok, Antarctica. \emph{Materialy Glyatsiol. Issled.} 91:129-133.

\bibitem[Lodders and Fegley(1998)]{1998psc..book.....L} Lodders, K., Fegley, B.\ (1998) \ The planetary scientist's companion / Katharina Lodders, Bruce Fegley.~ New York: Oxford University Press, 1998.~QB601.L84 1998.

\bibitem[Lunine and Stevenson(1985)]{1985ApJS...58..493L} Lunine, J.~I., Stevenson, D.~J. (1985) \ Thermodynamics of clathrate hydrate at low and high pressures with application to the outer solar system.\ \emph{The Astrophysical Journal Supplement Series} 58:493-531.

\bibitem[Luthi at al., 2010]{1ApJS...58..493L} Luthi, D., Bereiter, B., Stauffer, B., Winkler, R., Schwander, J., Kindler, P., Leuenberger, M., Kipfstuhl, S., Capron, E., Landais, A., Fisher, H., Stocker, T.F. (2010) \ CO$_2$ and O$_2$/N$_2$ variations in and just below the bubble-clathrate transformation zone of Antarctic ice cores. \emph{Earth and Planetary Science Letter} 297:226-233.

\bibitem[McKay et al.(2003)]{2003GeoRL..30m..35M} McKay, C.~P., Hand, K.~P., Doran, P.~T., Andersen, D.~T., Priscu, J.~C.\ (2003) \ Clathrate formation and the fate of noble and biologically useful gases in Lake Vostok, Antarctica.\ \emph{Geophysical Research Letters} 30:130000-1.

\bibitem[McKoy \& Sinano\u{g}lu (1963)]{McKoySinanoglu1961} McKoy, V., Sinano\u{g}lu, O., (1963) \ Theory of dissociation pressures of some gas hydrates.\ \emph{J. Chem. Phys.} 38 (12):2946-2956.

\bibitem[Miller(1974)]{1974PNAS...47.1798M} Miller, S.~L., (1974)\ The nature and occurrence of clathrate hydrates. In Natural Gases in Marine Sediments, edited by I. R. Kaplan, pp 151-177, Plenum Press, New York, 1974.

\bibitem[Miller(1961)]{1961PNAS...47.1798M} Miller, S.~L., (1961)\ The Occurrence of Gas Hydrates in the Solar System. \emph{Proceedings of the National Academy of Science} 47, 1798-1808.

\bibitem[Mohammadi et al.(2005)]{2005FaDi..147..509M} Mohammadi, A. H., Anderson, R., Tohidi, B. \ (2005) \ Carbon monoxide clathrate hydrates: equilibrium data and thermodynamic modeling.  \emph{AIChE Journal} 51:2825-2833.

\bibitem[Mousis et al.(2011)]{2011sf2a.conf..697M} Mousis, O., Waite, J.~H., Lunine, J.~I.\ (2011)\ Quantifying the measurement requirements needed to understand the origin of the Galilean satellite system.\ \emph{SF2A-2011: Proceedings of the Annual meeting of the French Society of Astronomy and Astrophysics} 697-699.

\bibitem[Mousis et al.(2010)]{2010FaDi..147..509M} Mousis, O., Lunine, J.~I., Picaud, S., Cordier, D.\ (2010) \ Volatile inventories in clathrate hydrates formed in the primordial nebula.\ \emph{Faraday Discussions} 147:509-525.

\bibitem[Mousis and Alibert(2006)]{2006A&A...448..771M} Mousis, O., Alibert, Y. (2006)\ Modeling the Jovian subnebula. II. Composition of regular satellite ices.\ \ \emph{Astronomy and Astrophysics} 448:771-778.

\bibitem[Mousis and Gautier(2004)]{2004P&SS...52..361M} Mousis, O., Gautier, D.\ (2004)\ Constraints on the presence of volatiles in Ganymede and Callisto from an evolutionary turbulent model of the Jovian subnebula. \ \emph{Planetary and Space Science} 52:361-370.

\bibitem[Parrish \& Prausnitz (1972)]{1972PP}Parrish, W.~R., Prausnitz, J.~M., (1972) \ Dissociation pressures of gas hydrates formed by gas mixtures.\ \emph{Industrial and Engineering Chemistry: Process Design and Development} 11 (1):26-35. Erratum : Parrish, W.~R., Prausnitz, J.~M., (1972) \ \emph{Industrial and Engineering Chemistry: Process Design and Development} 11 (3):462.

\bibitem[Pasek and Greenberg(2012)]{2012AsBio..12..151P} Pasek, M.~A., Greenberg, R.\ (2012) \ Acidification of Europa's Subsurface Ocean as a Consequence of Oxidant Delivery.\ \emph{Astrobiology} 12:151-159.

\bibitem[Philippe et al.(2001)]{2001Fi..147..509M} Philippe, J.-B., Petit, J.-R., Lipenkov, V. Ya., Raynaud, D., Barkov, N. I. (2001) \ Constraints on hydrothermal exchanges in Lake Vostok from helium isotopes.\ \emph{Nature} 411:460-462.

\bibitem[Priscu et al.(2001)]{2001F.147..509M} Priscu, J. C., Adams, E. E., Lyons, W. B., Voytek, M. A., Mogk, D. W., Brown, R. L., Mckay, C. P., Takacs, C. D., Welch, K. A., Wolf, C. F., Kirshtein, J. D., Avci, R.  (1999)  Geomicrobiology of subglacial ice above Lake Vostok, Antarctica.  \emph{Science} 286:2141-2144.

\bibitem[Redlich et al.(1949)]{1949F.147..509M} Redlich, O., Kwong, J. N. S. (1949) On the Thermodynamics of Solutions. V. An Equation of State. Fugacities of Gaseous Solutions. \emph{Chemical Reviews} 44 (1): 233-244.

\bibitem[Rudakov et (1949)]{194947..509M} Rudakov, E. S., Lutsyk, A. I., Mochalin V. N. (1996) Solubility of hydrocarbons in the system water--acetic acid. \emph{Ukrainian Chemistry Journal} 62 (9): 15-18.

\bibitem[Sander et al.(1949)]{1999F.147..509M} Sander, R. (1999) Compilation of Henry's Law constants for inorganic and organic species of potential importance in environmental chemistry (Version 3), http://henrys-law.org

\bibitem[Siegert et al.(2003)]{2003HyPr...17..795S} Siegert, M.~J., Tranter, M., Cynan Ellis-Evans, J., Priscu, J.~C., Berry Lyons, W.\ (2003) \ The hydrochemistry of Lake Vostok and the potential for life in Antarctic subglacial lakes.\ \emph{Hydrological Processes} 17:795-814.

\bibitem[Siegert et al.(2000)]{2000Natur.403..643S} Siegert, M.~J., Kwok, R., Mayer, C., Hubbard, B.\ (2000) \ Water exchange between the subglacial Lake Vostok and the overlying ice sheet.\ \emph{Nature} 403:643-646.

\bibitem[Showstack(2012)]{2012EOSTr..93R..80S} Showstack, R.\ (2012) \ Scientists Provide Perspectives as Drilling Reaches Subglacial Antarctic Lake Vostok.\ \emph{EOS Transactions} 93:80-81.

\bibitem[Sloan (2008)]{Sloan2008} Sloan, E. D. \& Koh, C. A., Clathrate Hydrates of Natural Gases, 3rd ed.; CRC Press, Taylor \& Francis Group, Boca Raton, 2008.

\bibitem[Smith et al.(2009)]{2009JGlac..55..573S} Smith, B.~E., Fricker, H.~A., Joughin, I.~R., Tulaczyk, S.\ (2009) \ An inventory of active subglacial lakes in Antarctica detected by ICESat (2003-2008).\ \emph{Journal of Glaciology} 55:573-595.

\bibitem[Thomas et al.(2009)]{2009P&SS...57...42T} Thomas, C., Mousis, O., Picaud, S., Ballenegger, V.\ (2009) \ Variability of the methane trapping in martian subsurface clathrate hydrates.\ \emph{Planetary and Space Science} 57:42-47.

\bibitem[Thomas et al.(2008)]{2008P&SS...56.1607T} Thomas, C., Picaud, S., Mousis, O., Ballenegger, V.\ (2008) \ A theoretical investigation into the trapping of noble gases by clathrates on Titan.\ \emph{Planetary and Space Science} 56:1607-1617.

\bibitem[Thomas et al.(2007)]{2007A&A...474L..17T} Thomas, C., Mousis, O., Ballenegger, V., Picaud, S.\ (2007) \ Clathrate hydrates as a sink of noble gases in Titan's atmosphere.\ \emph{Astronomy and Astrophysics} 474:L17-L20.

\bibitem[van de Flierdt et al.(2008)]{2008GeoRL..3521303V} van de Flierdt, T., Hemming, S.~R., Goldstein, S.~L., Gehrels, G.~E., Cox, S.~E.\ (2008) \ Evidence against a young volcanic origin of the Gamburtsev Subglacial Mountains, Antarctica.\ \emph{Geophysical Research Letters} 35:21303.

\bibitem[van der Waals \& Platteeuw (1959)]{vdWP1959} van der Waals, J.~H., Platteeuw, J.~C., 1959.\ Clathrate solutions. In: Advances in Chemical Physics, Vol. 2, Interscience, New York, 1-57.

\end{thebibliography}
\end{document}